\documentclass{Interspeech2024}
\usepackage{subcaption}

\interspeechcameraready

\title{A Multimodal Framework for the Assessment of the Schizophrenia Spectrum}

\name[affiliation={1}]{Gowtham}{Premananth}
\name[affiliation={1}]{Yashish M.}{Siriwardena}
\name[affiliation={1}]{Philip}{Resnik}
\name[affiliation={2}]{Sonia}{Bansal}
\name[affiliation={2}]{Deanna}{L.Kelly}
\name[affiliation={1}]{Carol}{Espy-Wilson}

\address{
  $^1$University of Maryland College Park, USA\\
  $^2$University of Maryland School of Medicine, USA }
\email{gowtham8@umd.edu, yashish@umd.edu, resnik@umd.edu, sbansal@som.umaryland.edu, dlkelly@som.umaryland.edu, espy@umd.edu }


\begin{document}

\maketitle

\begin{abstract}

This paper presents a novel multimodal framework to distinguish between different symptom classes of subjects in the schizophrenia spectrum and healthy controls using audio, video, and text modalities. We implemented Convolution Neural Network and Long Short Term Memory based unimodal models and experimented on various multimodal fusion approaches to come up with the proposed framework. We utilized a minimal Gated multimodal unit (mGMU) to obtain a bi-modal intermediate fusion of the features extracted from the input modalities before finally fusing the outputs of the bimodal fusions to perform subject-wise classifications. The use of mGMU units in the multimodal framework improved the performance in both weighted f1-score and weighted AUC-ROC scores.
    
\end{abstract}

\vspace*{-4pt}
\section{Introduction}

Schizophrenia is a highly debilitating and heterogeneous disorder manifested by positive (i.e. delusions, hallucinations, and disorganized thinking) and negative (i.e. diminished emotional expression, avolition, alogia, anhedonia, and asociality) symptoms. According to the World Health Organization's report, 1 in 300 people (0.32\%) worldwide is affected by schizophrenia \cite{institute2021global}. The Brief Psychiatric Rating Scale (BPRS) \cite{overall1962brief} is an assessment that clinicians or researchers use to measure psychiatric symptoms such as depression, anxiety, hallucinations, psychosis, and unusual behavior, which are categorized into 3 symptom categories Positive, Negative and Mixed symptoms \cite{10.1001/archpsyc.1982.04290070025006}.  The BPRS consists of 18 items which are scored in the range of 1-7 based on the severity of the symptom in the subject. The high level of heterogeneity in symptom presentation necessitates a multidimensional structure of symptomatology \cite{tonna2019dimensional}. While older studies pioneering subscale structures in assessment procedures like BPRS \cite{shafer2005meta} have provided a better understanding of symptom categories, considering the wide diversity in symptom presentation and phenomenology, in the vein of precision medicine, there is still a need to better identify clusters within symptoms based on measures beyond clinical ratings. 

Previous studies on multimodal approaches including various data modalities like audio, video, text, medical images, and medical records have proven to be successful in assisting clinicians in decision-making in different healthcare scenarios \cite{krones2024review}. This approach is particularly true for mental health disorders like major depressive disorder and schizophrenia \cite{CORCORAN2020770, Espy-Wilson2019,siriwardena2021multimodal,premananth2024multimodal}. Even though multiple studies have been conducted on schizophrenia, those studies only focused on a binary classification task where the systems developed classified the subjects into healthy controls and schizophrenia subjects. However, since the schizophrenia spectrum is so diverse, the classification has to be done in a multiclass manner so that the systems can differentiate between the different symptom categories.

Multimodal studies have been conducted in various domains \cite{krones2024review,arevalo2017gated,mGMU10234395} and different fusion techniques have been used in a lot of those studies. The way the modalities are combined has been instrumental in coming up with better predictive systems. In this work, we make use of the Gated Multimodal Units \cite{arevalo2017gated} which deploy a gate mechanism similar to that of logic gates to decide which modalities are fused and to what extent they impact the final results of the model. 

The multimodal frameworks that were developed for binary classification tasks in schizophrenia were not able to produce promising results in the multiclass classification tasks as the boundary between the symptom classes is not as definitive as the boundary between healthy controls and schizophrenia subjects. This motivated us to create a better machine-learning framework that could distinguish between symptom classes and healthy controls using unimodal and multimodal models. This work presents the importance of choosing the correct multimodal fusion technique and architecture when dealing with the task of differentiating between symptom categories of a neuropsychiatric disorder like schizophrenia.

\vspace*{-4pt}
\section{Dataset}

The dataset used for this study was collected for a mental health assessment project conducted at the University of Maryland School of Medicine in collaboration with the University of Maryland, College Park \cite{KELLY2020113496}. Interactive sessions were conducted where an investigator interviewed participants. The interviews progressed based on the responses of the participants. The sessions were videotaped and mental health assessments were made by trained psychiatrists before the interviews. Each subject was brought in for 1 to 5 sessions across 6 weeks and the mental health assessments were done before every session.

The dataset consists of data belonging to subjects with schizophrenia, depression, and healthy controls. For our study, we selected a subset (schizophrenia subset) of the data that included only the subjects who are on the schizophrenia spectrum and healthy controls. The schizophrenia subset includes multimodal data from 40 subjects with a total of 140 sessions. The data was classified into 3 classes namely healthy controls (HC), schizophrenia subjects with strong positive symptoms (P-SZ), and schizophrenia subjects with mixed symptoms (M-SZ). The subjects were assigned to these classes based on 2 criteria. First, the subjects were assigned to healthy controls and schizophrenia subjects based on the clinician's diagnosis. Then the subjects diagnosed with schizophrenia were assigned to the symptom classes based on their BPRS subscores. According to Shafer's work \cite{shafer2005meta}, the BPRS items unusual thought content, conceptual disorganization, hallucinatory behavior, and grandiosity are considered to be positive symptoms, and blunted affect, emotional withdrawal, and motor retardation are considered to be negative symptoms. The subjects who had an average score of 3.5 or higher only for the positive symptom items were assigned to the positive symptoms group. All the other subjects who were diagnosed with schizophrenia were assigned to the class of subjects with mixed symptoms as there were no subjects in the dataset who had an average score of 3.5 or higher for negative symptom items. The detailed description of the dataset and the class splits are provided in table \ref{tab:dataset}.

\begin{table}[h]
  \caption{\centering
  \textbf{Details of the Dataset used}(HC: Healthy Controls, M-SZ: Mixed symptoms, P-SZ: Positive Symptoms) }
  \label{tab:dataset}
  \centering
  \begin{tabular}{ l  c  c  c }
    \toprule
    {\textbf{}} & {\textbf{HC}}&{\textbf{M-SZ}}&{\textbf{P-SZ}} \\
    \midrule
    No.of Subjects&16&19&11\\
    No.of Sessions&54&56&30\\
    No of Utterances&7872&11574&7584\\
    Hours of Speech&9.82&12.16&12.47\\
    \bottomrule
  \end{tabular}
  
\end{table}

The dataset was manually divided into train-test-validation splits with (70:15:15) percent of the data. We considered each session as a separate sample because there were differences in the mental health state of the subjects based on the assessments reported by the physicians. However, all the sessions belonging to a subject were included in the same split so that the models trained would be speaker-independent and no bias would be introduced during the training or evaluation phases of the model.

\vspace*{-4pt}
\section{Feature extraction}

As the recordings consisted of speech from both the interviewer and the subject, the recordings from the interviews in the dataset were manually transcribed and diarized by a third-party transcription service. This allowed extraction of the subject's speech (from the interviewer) using the time stamps in the transcripts.
From the transcriptions obtained, the punctuation and stop words were first removed from the text as pre-processing measures before feature extraction. The text was tokenized into words using the NLTK (English) tokenizer. Finally, context-independent 100-dimensional GloVe word embeddings \cite{pennington2014glove} were extracted from the text. Then the extracted embeddings were stacked together as an array of width equal to the largest sentence available in the dataset and length equal to the maximum number of sentences available in the text of a session. The shorter sentences and the shorter sessions were padded to create equal-sized input features across all sessions.

Facial Action Units (FAUs) that track the movement and coordination of facial landmarks around the eyes, nose, and mouth regions of the subject during the speech were extracted from the video recordings using Openface~2.0: Facial Behaviour Analysis toolkit \cite{baltrusaitis2018openface}. Out of the extracted FAUs, 10 FAUs that mainly focus around the eyes and lip regions\cite{prince2015facial} were selected for the classification models as they have proven to be more effective in capturing changes manifested by mental health disorders \cite{siriwardena2021multimodal}.

Vocal tract variables (TVs) that track the changes in constriction degree and location of various articulators were the features extracted from the segmented audio recordings. The 6 TVs were derived from an acoustic-to-articulatory speech-inversion system \cite{attia2023improving}. In addition, an Aperiodicity, Periodicity, and Pitch detector \cite{deshmukh} was used to extract two glottal parameters; the amount of periodic and aperiodic energy in each frame. 

After extracting the low-level TVs and FAUs, high-level correlation structures were computed \cite{huang2020exploiting}. The Full Vocal Tract Coordination(FVTC) matrix is produced by calculating delayed autocorrelation and cross-correlation across the low-level features starting from 0 to a delay of 'D' frames. For TVs, the best performance was obtained with D=50 while for FAUs, the best performance was obtained with D=45 \cite{siriwardena2021multimodal}.

\vspace*{-4pt}
\section{Methodology}

\begin{figure*}[th!]
  \centering
  \includegraphics[width=\textwidth, height=5 cm]{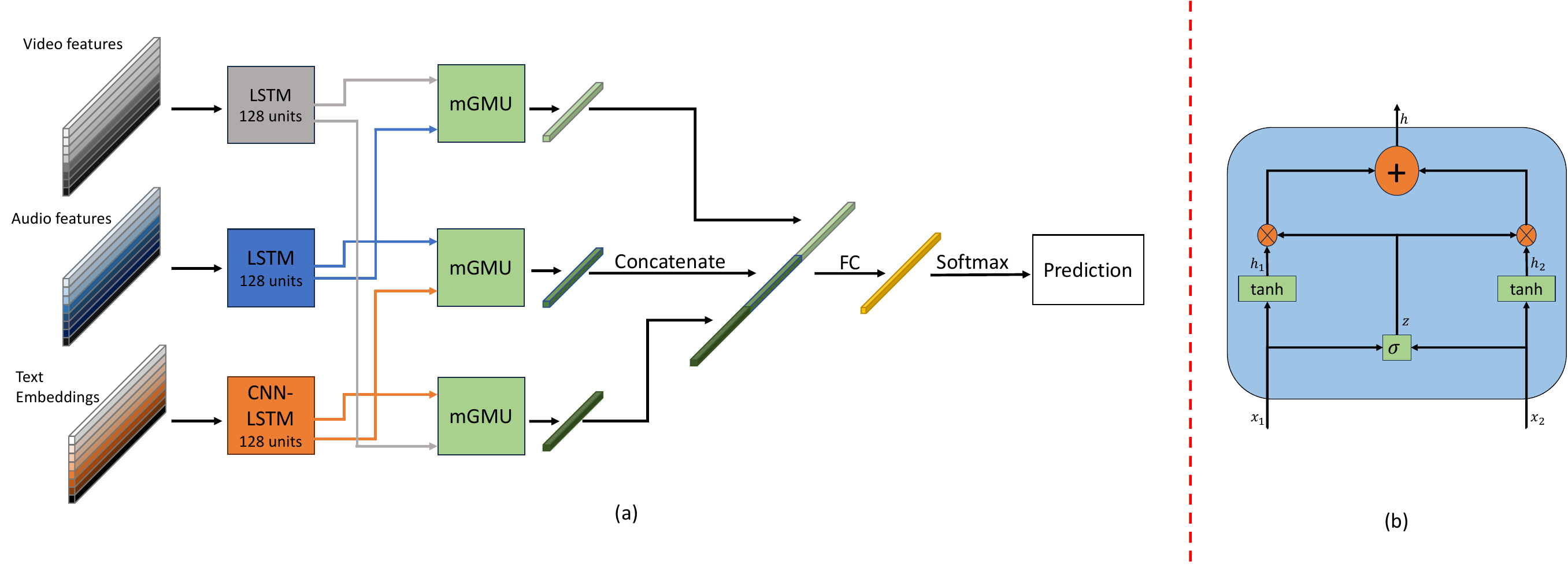}
  \caption{(a): Multimodal Framework with mGMU (b): Schematic diagram of Minimal Gated Multimodal Unit.}
  \label{fig:mm_GMU}
\end{figure*}

\subsection{Fusion methods}

The fusion approaches used to combine multimodal data can be broadly categorized into three: \textbf{Early Fusion (E-F)} which is the fusion of the original feature representation of the modalities before they are sent through any models; \textbf{Intermediate Fusion (I-F)} which fuses the latent space representations of the modalities to create a combined representation where the extracted features of the individual modalities have been processed individually through separate models to a certain extent but not up to the final decision; and \textbf{Late Fusion (L-F)} where the fusion happens when all the modalities are processed separately their individual predictions are fused to get a final prediction. Some recent studies have also used hybrid approaches \cite{bridgetower} where they have utilized more than one of those approaches. 
Regardless of the fusion approaches, various fusion methods have been deployed to integrate the feature vectors or the predictions of the individual modalities. The commonly used fusion methods are,
\begin{enumerate}
\item \textbf{Concatenation}: Feature vectors are concatenated directly to create a longer feature vector
\item \textbf{Operation-based}: Feature vectors are fused by performing element-wise operations like addition or multiplication of the feature vectors
\item \textbf{Learning-based}: Feature vectors are combined using specialized machine learning models to create the combined representations.
\end{enumerate}
In this study, we tried using various fusion methods and fusion approaches to come up with a multimodal framework that can classify subjects on the schizophrenia spectrum from healthy controls. 

\subsection{Minimal Gated Multimodal Unit}

A Gated multimodal unit (GMU) \cite{arevalo2017gated} is a neural network component used for multimodal data fusion which gets the latent space representation of different modalities and produces a combined representation based on a gate neuron that oversees and adjusts the contribution of each modality's representation in the output of the unit. The minimal gated multimodal unit (mGMU) \cite{mGMU10234395} which uses only one gate, is a variant of the GMU that was introduced to reduce the computational complexity of the GMU but was still able to produce similar results. In this study, we made use of the mGMU as a part of our architecture for multimodal fusion. fig.\ref{fig:mm_GMU}.(b) shows the architecture of the mGMU where $x_1$ and $x_2$ are input feature representations and $h$ is the produced combined representation. The equations governing the function of the mGMU are as follows:  
\begin{align} 
h_1 = tanh(W_1 \cdot x_1) \nonumber\\
h_2 = tanh(W_2 \cdot x_2)\\
z = \sigma (W_z \cdot [x_1,x_2])\\
h = z * h_1 + z * h_2\\
\theta = \{W_1, W_2, W_z\} \label{parameter equation}
\end{align}

The $\theta$ in equation \ref{parameter equation} is the set of parameters that will be trained during the training stage to create combined representations of the features.

\subsection{Uni-modal model architectures} 

\begin{figure}[h!]
  \centering
  \includegraphics[width=\linewidth]{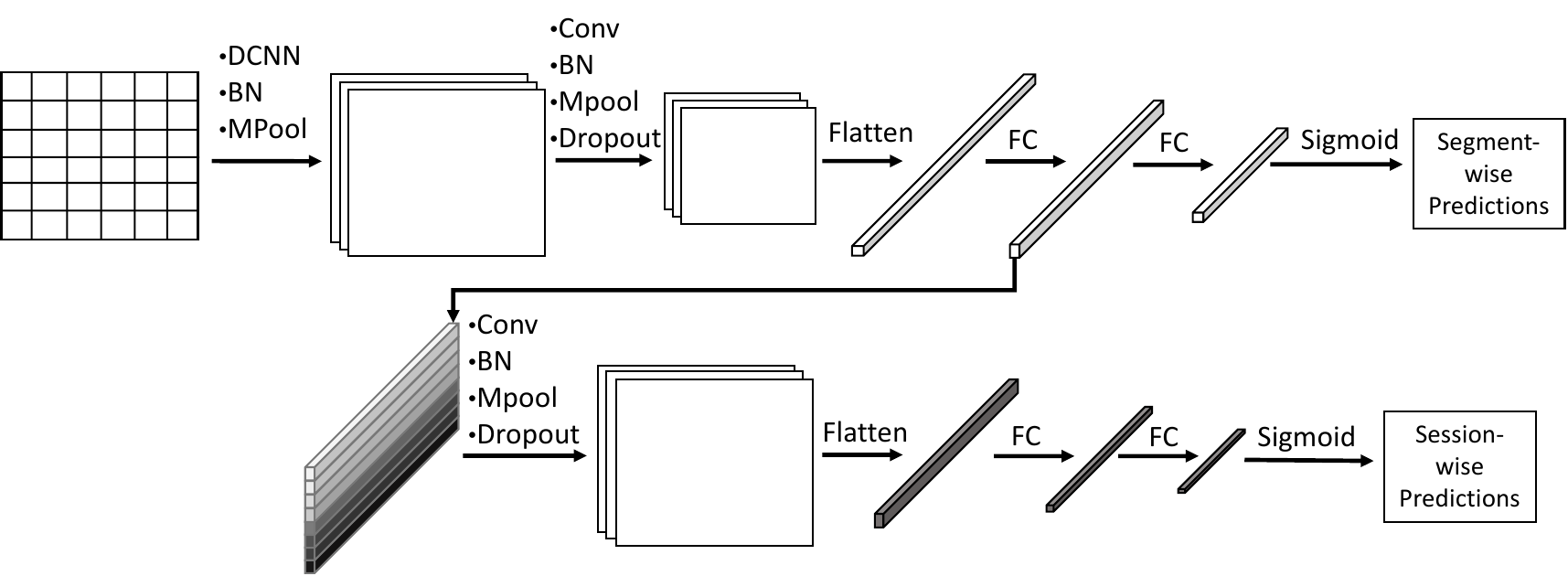}
  \caption{Unimodal architecture used for the audio and video modalities.}
  \label{fig:video}
\end{figure} 

Fig.\ref{fig:video} shows the unimodal model implemented for the audio and video modalities. It is a segment-to-session-level classification neural network (STS-CNN) inspired by the work of Premananth et al.\cite{premananth2024multimodal}. Dilated Convolution Neural Network (DCNN) \cite{huang2020exploiting} with dilation rates of 1,3,7,15 was used for the segment-level classifier and a simple Convolution Neural Network (CNN) for the session-level classifier. The FVTC correlation structures were used as inputs for the segment-level classifiers. The intermediate outputs of vectors of length 128 were obtained from the segment-level classifier and the vectors of all the segments belonging to a session were stacked and used as input feature vectors for the session-level classifiers.  

\begin{figure}[h!]
  \centering
  \includegraphics[width=\linewidth]{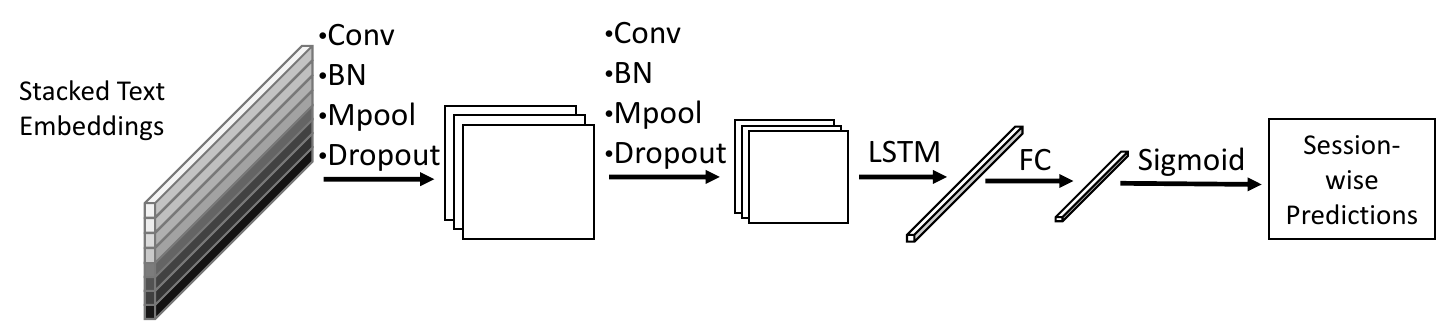}
  \caption{Unimodal architecture used for the text modality.}
  \label{fig:text}
\end{figure}

A Convolution Neural Network-Long Short Term Memory (CNN-LSTM) based model architecture as shown in fig.\ref{fig:text} was implemented for the text-based unimodal system.
Glove embeddings extracted from the text of the subject were used as inputs for the model.

\subsection{Multimodal model architecture} 
\begin{table*}[th!]
  \caption{\centering 
  \textbf{Summary of the classification models} (I-F: Intermediate Fusion, CI: Confidence Interval)}
  \label{tab:class_models}
  \centering
  \begin{tabular}{ c  c  c  c  c  c  c}
    \toprule
    {\textbf{Modality}} & {\textbf{Model}}&{\textbf{Fusion}}&\textbf{Weighted F1}& \textbf{95\% CI for F1}&{\textbf{AUC-ROC}}& \textbf{Confusion matrix} \\
    \midrule
    Audio&STS-CNN \cite{premananth2024multimodal}&-&0.5695&[0.3827,0.7791]&0.8069&[[6,2,0],[2,5,2],[0,4,2]] \\
    Audio&STS-CNN (ours)&-&0.6122&[0.4012,0.7926]&0.8157&[[6,2,0],[2,5,2],[1,2,3]] \\
    Video&STS-CNN \cite{premananth2024multimodal}&-&0.5368&[0.3281,0.7469]&0.7032&[[6,2,0],[3,3,3],[2,1,3]]\\
    Video&STS-CNN (ours)&-&0.5395&[0.3289,0.7343]&0.7435&[[7,1,0],[5,1,3],[2,1,3]]\\
    Text&HAN \cite{premananth2024multimodal}&-&0.5739&[0.3528,0.7560]&0.5794&[[7,0,1],[5,0,4],[0,2,4]]\\
    Text&CNN-LSTM (ours)&-&0.5915&[0.3965,0.7936]&0.7846&[[2,6,0],[0,6,3],[0,1,5]]\\
    \midrule
    Multimodal&CNN-HAN \cite{premananth2024multimodal}&I-F with Attention&0.5162&[0.2963,0.6587]&0.7421&[[2,6,0],[1,6,2],[0,1,5]]\\
    \textbf{Multimodal}&\textbf{CNN-LSTM (ours)} &\textbf{I-F with mGMU}&\textbf{0.6547}&\textbf{[0.4569,0.8396]}&\textbf{0.8214}&\textbf{[[6,2,0],[2,5,2],[1,1,4]]}\\
    \bottomrule
  \end{tabular}  
\end{table*}

A Learning-based intermediate fusion approach was used for the multimodal architecture as shown in fig.\ref{fig:mm_GMU}.(a) was implemented in this study. This model takes the stacked session-level audio and video features obtained from the unimodal models and the GloVe embeddings of the transcribed text as the input. The audio and video features are sent through a pair of LSTM layers while the text embeddings are sent through a set of CNN layers followed by an LSTM layer. The latent space representations obtained after the three LSTM layers of all three modalities are used as inputs to 3 mGMU units which produce three bimodal combined representations (audio-video, audio-text, video-text). The combined representations are concatenated together and sent through fully connected layers to produce the final classification labels.

\vspace*{-4pt}
\section{Experiments and Results}

The longer segments in the data were segmented into 20-second segments with an overlap of 5 seconds for the STS-CNN video model and segmented into 40-second segments with an overlap of 5 seconds for the STS-CNN audio model. The segment lengths were chosen based on a grid search done to find the optimum segment length from a set of (20,30,40) seconds.

A grid search was done for the initial hyperparameter tuning for the models. Initial Learning rate, learning rate patience, early stopping patience, and learning rate reduction factor were chosen using grid search from sets of (1e-3,5e-4,1e-4), (20,25,30), (40,50,60), and (0.75,0.5,0.25). Based on the results obtained from the grid search, we trained all the models for 300 Epochs with an ADAM optimizer with an initial learning rate of 1e-3 with a learning rate patience of 25 epochs. If there was no improvement in the validation loss for 25 epochs, the learning rate was reduced by a factor of 0.5. The models were also trained using early stopping criteria for monitoring the validation loss improvement for 50 epochs. During the model training, class weights were applied to the loss function of the models because of the class imbalance found in the dataset to reduce the model being over-fitted to the class with more samples. All the models were implemented using a Tensorflow-Keras framework. The proposed multimodal framework contains 897k trainable parameters. The multimodal model takes 12 minutes and 30 seconds on average to train on a GeForce RTX 3090 GPU when all the input features are already extracted and preprocessed by the unimodal models.  

The models were evaluated based on the weighted F1-score with a 95\% confidence interval and weighted Area under the Curve (AUC) Receiver Operating Characteristic(ROC) curve. The confidence intervals were calculated using the software developed by Ferrer, L. and Riera, P. \cite{conf}. The AUC-ROC score is calculated by getting the area under the ROC curve which is obtained based on the behavior of the classifier for every threshold by plotting the True positive rate (TPR) and False positive rate (FPR). As this is a multi-class classification problem, the `one vs rest' configuration of the AUC-ROC score was used which considers the class for which it is being calculated as the positive class and all the other classes combined as the negative class. The per-class AUC-ROC scores were calculated and they were averaged using a weighted average that considers the number of samples available for each class in the test split.

Table \ref{tab:class_models} compares the performance of the proposed best-performing unimodal and multimodal models in this study with previous state-of-the-art models \cite{premananth2024multimodal}. The comparison of the unimodal models in this study was constrained to only the best-performing models from a previous study because the comparison with other baselines has been done in our previous works \cite{siriwardena2021multimodal,premananth2024multimodal}. The confusion matrices shown in the table are 3x3 matrices where the rows of the matrix represent the true labels and columns represent the predicted labels in the order of HC, M-SZ, P-SZ. These confusion matrices correspond to the evaluation results of the models on the Test-split containing a total of 23 sessions. An ablation study was carried out on the proposed multimodal framework to validate the components included in the framework and also to validate the data fusion approaches used in the framework. The results of the ablation study are tabulated in Table \ref{tab:ablation}.
\vspace{-5pt}
\begin{table}[h!]
  \caption{\centering
  \textbf{Summary of the Ablation study} (I-F: Intermediate Fusion, L-F: Late Fusion)}
  \label{tab:ablation}
  \centering
  \begin{tabular}{  l  c  c}
    \toprule
    {\textbf{Fusion approach \& method}}&{\textbf{Weighted F1}}&{\textbf{AUC-ROC}} \\
    \midrule
    L-F without mGMU&0.4808&0.8127\\
    L-F with mGMU&0.5560&0.6830\\
    I-F without mGMU&0.5538&0.7859\\
    \textbf{I-F with mGMU}&\textbf{0.6547}&\textbf{0.8214}\\
    \bottomrule
    \end{tabular}
  
\end{table}

\vspace*{-4pt}
\section{Discussion}

This study was focused on the 3-class classification task of classifying subjects into healthy controls, subjects with mixed schizophrenia symptoms, and subjects with positive schizophrenia symptoms. As previous studies have only focused on a binary classification between subjects with and without schizophrenia, our study had a higher level of difficulty as the boundaries between different symptom classes are not that rigid. There could be subjects who have high severity scores for positive symptoms and high severity scores for negative symptoms and should be diagnosed as subjects with mixed symptoms also there could be subjects with low severity scores in both positive and negative symptoms and will still have to be diagnosed as mixed symptoms. This uncertainty between the class boundaries required us to create a multimodal framework that should capture various traits from the audio, video, and text modalities together to come up with a reliable classification system. 

One of the key observations from this study is that all the unimodal and multimodal systems proposed in this study have outperformed previous state-of-the-art models used for distinguishing schizophrenia subjects from healthy controls. Another significant improvement made with this framework is that it uses only 897k trainable parameters while the previous state-of-the-art model \cite{premananth2024multimodal} uses around 1930k trainable parameters.

The results from the ablation study performed to validate the multimodal framework tabulated in Table \ref{tab:ablation} show that for this multiclass classification problem, the selection of Intermediate-Fusing with the use of minimal Gated multimodal units was the best choice as all the other combinations of fusion approaches and methods were not able to match or better the performance of the proposed framework. The early Fusion approach was not tested as the different modalities had initial feature vectors in different shapes and therefore a straightforward early feature fusion was not possible.

\vspace*{-4pt}
\section{Conclusion and Future work}

In conclusion, this paper opens a new avenue of classification of subjects in the schizophrenia spectrum based on their symptom classes. A novel multimodal architecture is proposed that makes use of mGMU to classify the subjects in the schizophrenia spectrum from healthy controls. As of now, due to data limitations, we were only able to test the model on 3 classes barring the schizophrenia subjects with strong negative symptoms. In future work, we would like to expand our system to be a 4-class classification that can clearly distinguish all 3 classes of schizophrenia subjects and healthy controls. Additionally, we would like to build systems that can predict the severity scores of the symptom classes or sub-categories which will provide the clinicians with better insights to structure their treatment procedures.

\section{Acknowledgement}

This work was supported by the National Science Foundation grant numbered 2124270
.
\bibliographystyle{IEEEtran}
\bibliography{paper}

\end{document}